# Light–induced magnetization reversal of high-anisotropy TbCo alloy films

Sabine Alebrand[1], Matthias Gottwald[2,3], Michel Hehn[2], Daniel Steil[1], Mirko Cinchetti[1], Daniel Lacour[2], Eric E. Fullerton[3], Martin Aeschlimann[1], Stéphane Mangin*[2]

[1]*Department of Physics and Research Center OPTIMAS, University of Kaiserslautern, D-67663 Kaiserslautern, Germany*

[2]*Institut Jean Lamour UMR CNRS 7198– Université de Lorraine, Vandoeuvre-lès-Nancy, France*

[3]*Center of Magnetic Recording Research, University of California San Diego, CA, USA*

* for more information contact: stephane.mangin@ijl.nancy-universite.fr

**Magnetization reversal using circularly polarized light provides a new way to control magnetization without any external magnetic field and has the potential to revolutionize magnetic data storage. However, in order to reach ultra-high density data storage, high anisotropy media providing thermal stability are needed. Here, we evidence all-optical magnetization switching for different $Tb_xCo_{1-x}$ ferrimagnetic alloy composition and demonstrate all-optical switching for films with anisotropy fields reaching 6 T corresponding to anisotropy constants of $3\times10^6$ ergs/cm$^3$. Optical magnetization switching is observed only for alloys which compensation temperature can be reached through sample heating.**

Magnetization and spin manipulation using excitations such as short-pulse or rf magnetic fields [3], electric fields [4], temperature pulses [5], polarized currents [6] or polarized light [1] is a current topic of great interest, particularly for magnetic data storage application. It is generally believed that to reach ultra-high densities (above 1 Tb/inch$^2$) the hard-disk-drive industry will need a transition to new technologies, such as bit-patterned or heat-assisted magnetic recording [7-9]. The reason is that as the bit size is reduced, recording materials with small grains and strong perpendicular magnetic anisotropies (PMA) are needed to maintain signal to noise ratios and thermal stability for archival storage. However, such large anisotropies require magnetic fields much larger than those generated by a write head. This competition between thermal stability and addressability is known as the superparamagnetic limit [10].

Heat-assisted magnetic recording addresses the superparamagnetic limit by using laser light in order to heat the medium during the writing process and to consequently decrease the switching field down to reasonable values [8]. However, this requires integrated magnetic field and plasmonic optical write elements [9]. An intriguing pathway is to directly switch magnetic domains using circularly polarized light without any applied magnetic field. In a pioneering work the T. Rasing group in Nijmegen showed in 2007 fully deterministic magnetization switching in a ferrimagnetic GdFeCo alloy film using single circularly polarized femtosecond laser pulses [1]. This remarkable phenomenon has been termed all-optical switching (AOS) and occurs on a timescale of a few tens of picoseconds [11], about two orders of magnitude faster than anything feasible without using relativistic electron sources [12]. Thus, AOS may lead to technological breakthroughs by using polarized light for ultrafast magnetization manipulation for a variety of applications. As such, considerable recent effort has been made to investigate the physics of the AOS and revealing the underlying microscopic mechanism [2, 11, 14,16].

Despite the recent progress in the understanding of AOS, many fundamental questions still need to be addressed. AOS has been observed in GdFeCo alloy films and it is still unclear how important the specific material properties are. In amorphous GdFeCo ferrimagnetic rare earth-transition metal alloys the magnetic moment carried by the rare earth (Gd) is antiparallel to the moment held by the transition metal (FeCo) sub-lattice. Depending on the concentration there exists a compensation temperature ($T_{Comp}$) at which the net magnetization, resulting from the two sub-lattices, vanishes [17]. Since in GdFeCo AOS is observed in samples with $T_{Comp}$ both above and below room temperature [2], it seems that AOS proceeds independently from the presence of a compensation point.

A unique feature of GdFeCo alloys is that Gd is an S-ion without any orbital momentum contribution to its magnetization. This contributes to the rather weak perpendicular magneto-crystalline anisotropy (PMA) observed in GdFeCo films that considerably limits their relevance for potential application. Crucially, for ultra-high density data storage optically recordable magnetic materials exhibiting larger PMA are needed. Staying in the class of rare earth-transition metal alloys the most straightforward solution would be to replace Gd with Tb, since Tb has a significant orbital momentum. Indeed ferrimagnetic $Tb_xCo_{1-x}$ alloy films exhibit strong PMA and have been already used in conventional magneto-optical recording [18].

Here we study the feasibility of all-optical switching on $Tb_xCo_{1-x}$ films which exhibit a strong perpendicular magnetic anisotropy [19,20], varying the composition from x=12% to x=34%. We demonstrate that the observation of all-optical switching depends on both the Tb concentration and the properties of the exiting laser pulse.

To detect the AOS we used a Faraday imaging setup consisting of a white-light source, a crossed polarizer pair and a CCD-camera, as schematically depicted in Figure 1a. Consequently, opposite directions of the sample magnetization can be seen as black and white contrast, as exemplarily shown in Figure 1b. To finally switch the magnetization we utilized either a ps-laser running at 5 kHz at a central wavelength of 532 nm (FWHM at sample position ~10 ps) or a fs-amplifier running at 6.3 kHz at a central wavelength of 780 nm (FWHM at sample position ~400 fs). The pulses were circularly polarized by a zero order quarter-wave plate and focused on the sample.

During the experiments the ultrashort laser is swept over a defined and homogeneously magnetized area of the sample and the occurrence of AOS is monitored by the Faraday imaging setup. As AOS only works in a relatively narrow fluence range a threshold value has to be overcome to obtain AOS [11]. In case of too high fluences pure thermal demagnetization occurs. In that case one obtains helicity-independent switching of the magnetization, resulting in a multidomain state. In the gray-scale Faraday images this shows up as neither black nor white but as a gray contrast. The minimum threshold fluence above which switching occurs was monitored. To determine the latter the laser intensity was continuously varied until the threshold from "no switching" to either "all-optical switching" or "pure thermal demagnetization" was found. To distinguish between all-optical switching and thermal demagnetization the switching behavior was checked for the different combinations of laser helicity and sample magnetization [14]. If switching occurred independently of the combination used it was attributed as pure thermal demagnetization. All measurements were performed at room temperature.

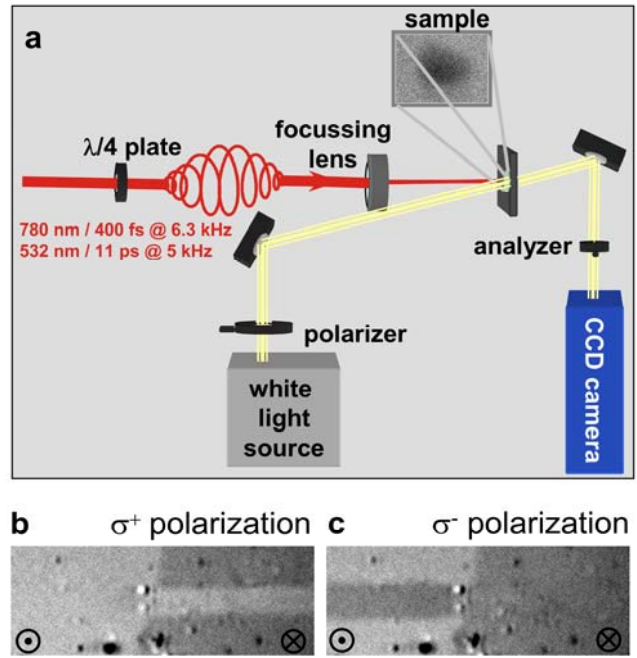

*Figure 1: All-optical switching experiment. (a): Schematic of the AOS setup. (b) and (c): Demonstration of AOS of a $Tb_{26}Co_{74}$ film. Two out-of-plane oriented domains of opposite direction were created by an external magnetic field, showing up as black and white contrast. Then the laser pulse with fixed circular polarization was swept over the sample and a stripe domain was written (b). Afterwards the circular polarization was inverted and the before written domain could be deleted while a stripe domain of opposite orientation appeared (c). The images were processed to optimize the contrast between domains of opposite orientation.*

The TbCo films are multilayer samples of Glass//Ta(5 nm) / $Tb_xCo_{1-x}$ (20 nm) / Cu(2 nm) / Pt(5 nm) grown by DC magnetron sputtering. The base pressure during sputtering was less than $5\times10^{-9}$ mbar to ensure oxygen free layers. Pure Co and Tb targets were used for a co-sputtering process and the relative atomic concentration of the two elements Co and Tb was controlled by the relative sputtering power. For all compositions (x=12% to x= 34%.) the magnetization lies perpendicular to the film plane. We characterized the samples in terms of compensation temperature, saturation magnetization and anisotropy fields. Below 400 K these quantities were measured using a Quantum Design PPMS system, with precisely controlled temperature under He atmosphere. Measurements above 400 K were done using a Lakeshore VSM. Temperature was controlled via the flow of hot Ar gas allowing less accuracy than for the low temperature measurements.

In Figure 2 the room temperature saturation magnetization ($M_s$) and anisotropy field ($H_K$) as a function of the sample composition is presented. The measured anisotropy field ranging up to 6 Tesla and corresponding to an anisotropy constant of $3.10^6$ ergs/cm$^3$ is compatible with high thermal stability for high-density patterned recording media.

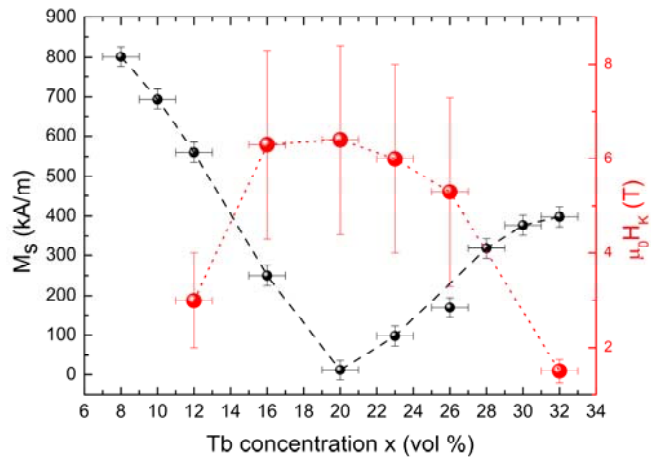

*Figure 2: Magnetic properties of the $Tb_xCo_{1-x}$ samples. Saturation magnetization $M_S$ (black circles) and anisotropy field $H_K$ (red circles) for the different $Tb_xCo_{1-x}$ samples. The dashed and dotted lines are interpolations between the measured points. Large error bars are due to the uncertainty of magnetometry measurements for thin film samples with small magnetization.*

Figure 3 shows the measured $T_{Comp}$ and the Curie temperatures ($T_{Curie}$) as a function of the Tb concentration. While $T_{Comp}$ increases with increasing Tb concentration (x), $T_{Curie}$ decreases. Depending on the position of $T_{Comp}$ and $T_{Curie}$ relative to room temperature ($T_{RT}$) one can distinguish three regions: Region I is defined as the one for which $T_{Comp} < T_{RT} < T_{Curie}$ corresponding to x< 22%. As the Tb concentration increases, $T_{Comp}$ increases and region II appears where $T_{RT} < T_{Comp} < T_{Curie}$ for 22% <x< 30%. For samples with x> 28%, $T_{Comp}$ and $T_{Curie}$ are quite close. Region III is defined as the one for which x> 30%, where the compensation temperature no longer exists.

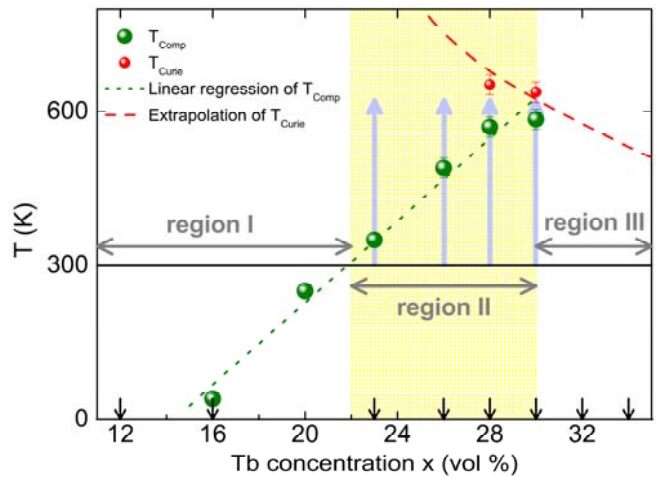

*Figure 3: Compensation temperature $T_{Comp}$ and Curie temperature $T_{Curie}$ dependence on the composition. $T_{Comp}$ is shown as green large circles, $T_{Curie}$ as red small circles. The green dotted line is guide to the eye for $T_{Comp}$. The red dashed line corresponds to published results of $T_{Curie}$ [17], whereby the curve was vertically shifted to match our data points. The small black arrows mark the compositions used for magnetization switching experiments on the x-axis. The horizontal solid black line marks room temperature. The blue-gray arrows are schematically showing that, as the laser heats the sample, its temperature increases and eventually crosses the compensation temperature.*

For our AOS measurements, we used samples from all three regions, having: x=12%, 16% (region I), x=23, 26, 28, 30% (region II) and x=32%, 34% (region III). We stress that all these TbCo samples exhibit a very high anisotropy field, up to 6 Tesla for x= 23% (corresponding to an anisotropy constant of $3 \times 10^6$ ergs/cm$^3$) (see Fig. 2). At room-temperature we observe AOS exclusively for samples in region II as shown in Figs. 2(b) and 2(c) for the $Tb_{26}Co_{74}$ film where one circular light polarization reverses the magnetization in one direction whereas inversing the circular polarization reverses the magnetization in the opposite direction. A real-time video of the magnetization switching is available on http:// youtube. The minimum fluence needed to observe AOS is in the order of a few mJ/cm$^2$. By further increasing the fluence pure thermal demagnetization occurs. For samples in region I and III only pure thermal demagnetization is observed and the particular circular polarization has no effect on the magnetization switching.

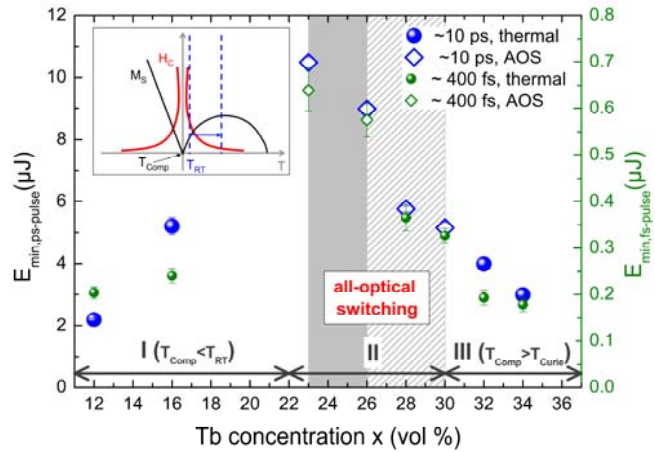

*Figure 4: Results of the all-optical switching experiments.* *Minimum laser pulse energy $E_{min,pulse}$ required either for AOS (open diamonds) or pure thermal demagnetization (full circles), depending if AOS works or not. The results are shown for fs (small symbols) and ps (large symbols) pulse duration of the exciting laser pulse. The gray background marks the*

*range where all-optical switching is obtained for both ps and fs pulses while in the range of the gray shaded background all-optical switching was only observed for ps pulses. For all other compositions no AOS but pure thermal demagnetization was observed. The error bars result from the laser power fluctuations. The inset shows the typical behavior of the saturation magnetization and coercivity around the compensation point.*

In Fig. 4 we plot the minimum laser pulse energy $E_{min,pulse}$ either for AOS (open symbols) or pure thermal demagnetization (full symbols) [[24] The qualitative behavior remains the same if one considers fluences as one only divides by a constant factor (spot size) to obtain the fluence from the pulse energy. In our discussion we therefore restrict ourselves on discussing trends instead of absolute values.]. The most striking feature is that AOS is only observed for samples with $T_{Comp}$ above room temperature but below $T_{Curie}$. Furthermore for fs pulses only the $Tb_{23}Co_{77}$ and $Tb_{26}Co_{74}$ samples of region II could be switched all-optically (marked by the gray background in Fig.4), while for picosecond pulses AOS *additionally* works for the samples with Tb concentration x=28 % and x=30 % (marked by the shaded background in Fig.4). We expect this to be caused by the different pulse duration but it should also be mentioned that the central wavelength of the exciting laser pulse varied for the two measurements (780 nm for fs pulses and 532 nm for ps pulses). Additionally we report that also the samples which show AOS seem to be slightly sensitive to thermal demagnetization effects: Although we find a clear difference for the switching behavior on the laser helicity (attributed thus to AOS), often a slight thermal contribution appears at the threshold for AOS.

The above observations are not only relevant for the microcopic understanding of AOS but for technological purpose as well.

Fig. 3 indicates that the crossing of $T_{Comp}$ after laser excitation is crucial to enable AOS in this material system. Indeed for samples in region I, $T_{Comp}$ is below room temperature and, hence, it cannot be crossed due to laser heating when performing the experiments at room temperature. For the $Tb_xCo_{1-x}$ thin films in region II, compensation temperature is always above room temperature, consequently laser heating helps crossing the compensation temperature and reversal occurs (marked by the blue-gray arrows in Fig.3). For region III $T_{Comp}$ no longer exists and therefore cannot be crossed either. In accordance AOS is not observed.

We emphasize that the importance of $T_{Comp}$ for the feasibility of AOS is quite different to what has been found for GdFeCo alloys. For the latter all-optical switching is reported to take place above and below the compensation point [24]. Moreover in GdFeCo the magnetization switching would be taking place because of the magnetic field generated by light [24]. We then wonder if the same mechanism can explain our observation since the low magnetization and the high anisotropy of the considered TbCo induce very small coupling with any magnetic field.

On the other hand since the crossing of compensation temperature seems to be required for AOS process the importance of angular momentum transfer could be revisited [21]. Indeed, as it has been clearly pointed out that the amount of angular momentum delivered by the light is extremely small, one can argue that at compensation very little momentum is require for switching.

The evolution of minimum laser pulse energy $E_{min,pulse}$ needed either for AOS or pure thermal demagnetization shown on figure 4 raise several questions. Why is the minimum laser energy needed for AOS larger than for thermal demagnetization? Why does it decreases for AOS as we move away from compensation by increasing Tb Composition in the alloys? Do the specific material properties (e.g. coercivity) which are related to the position of the compensation temperature relative to room temperature (see inset in Fig. 4) influence the observability of AOS? To our knowledge no theory permits to fully explain those data which underline that a better understanding of the interaction between polarized light and magnetization is still needed.

In conclusion we have shown all-optical magnetization switching in $Tb_xCo_{1-x}$ ferrimagnetic alloy films for a certain composition range. This composition range corresponds to the particular one for which the compensation temperature is between room temperature and the Curie temperature. Switching has been demonstrated for samples exhibiting anisotropy fields up to 6 T and anisotropy constants of $3x10^6$ ergs/cm$^3$. These experiments demonstrate the potential of the use of polarized light for magnetic data storage technology and provide a crucial piece of information for understanding the physics of all-optical switching.

## Acknowledgements

We acknowledge the Photonik-Zentrum Kaiserslautern e.V., especially Dr. J. L'huillier for access to and T. Herrmann and M. Stolze for technical support with the ps laser system. Further we thank A. Hassdenteufel for establishing the switching experiment. We thank Region Lorraine, Institut Universitaire de France, Grosse/Grande Region for their financial support. We thank The CC-Magnetisme from IJL and especially T. Hauet and S. Suire for their support with magnetic measurements. This work was also supported by The Partner University Fund "Novel Magnetic Materials for Spin Torque Physics and Devices" and the ANR-10-BLAN-1005 „Friends" and ANR-09-Blan-076 „SpinPress"

## References

[1] Stanciu**,** C.D, Hansteen**,** F., Kimel**,** A.V. , Kirilyuk**,** A., Tsukamoto**,** A., Itoh**,** A. & Rasing**,** Th**.** Phys. Rev. Lett. 99, 047601 (2007)

[2] Vahaplar, K., Kalashnikova, A. M. , Kimel, A. V. , Gerlach, S., Hinzke, D. , Nowak, U., Chantrell, R. , Tsukamoto, A. , Itoh, A. , Kirilyuk, A. & Rasing, Th.
Phys. Rev. B 85, 104402 (2012)

[3] Van Waeyenberge, B., Puzic, A. , Stoll, H. , Chou, K. W. , Tyliszczak, T. , Hertel, R. Fähnle, M. , Brückl, H. , Rott, K. , Reiss, G. , Neudecker, I. , Weiss, D. , Back, C. H. & Schütz G.,. Nature 444, 461-464 (2006)

[4] Shiota, T., Nozaki, T. , Bonell, F. , Murakami, S. , Shinjo, T. , Suzuki, Y. Nature Materials 11, 39 (2012)

[5] Slonczewski, J. C.. Phys. Rev. B 82, 054403 (2010)

[6] Mangin, S. , Ravelosona , D. , Katine, J. A. , Carey, M. J. , Terris, B. D. & Fullerton, E. Nature Materials 5, 210 - 215 (2006)

[7] Ozatay, P. G. Mather, J. U. Thiele, T. Hauet, and P. M. Braganca, "Spin-Based Data Storage", in Nanofabrication and Devices, Vol. 4, of Comprehensive Nanoscience and Nanotechnology, D. Andrews, G. Scholes, G. Wiederrecht, eds. (Elsevier, London), 2010; 561-614.

[8] Kryder, M. H. , Gage, E. C. , McDaniel, T. W. , Challener, W. A.,
Rottmayer, R. E. , Ju, G. , Hsia, Y.-T. & Fatih Erden, M. Proceedings of the IEEE Vol. 96, No. 11, November 2008, 1810-1835.

[9] Stipe, B.C., Strand, T.C., Poon, C.C., Balamane, H., Boone, T.D., Katine, J.A., Li, J.-L., Rawat, V., Nemoto, H., Hirotsune, A., Hellwig, O., Ruiz, R., Dobisz, E.,  Kercher, D.S., Robertson, N., Albrecht T.R. & Terris. B.D. Nature Photonics, Volume: 4, Pages: 484-488 (2010)

[10]  Moser, A., Takano, K., Margulies, D. T., Albrecht, M. , Yoshiaki, S., Ikeda, Y. , Sun, S. and Fullerton, E. E. Magnetic, J. Phys. D: Appl. Phys. 35, R157 (2002).

[11] Vahaplar, K., Kalashnikova, A. M. , Kimel, A. V. , Hinzke, D. ,  Nowak, U. , Chantrell, R. , Tsukamoto, A. , Itoh, A. , Kirilyuk, A.  & Rasing, Th.. Phys. Rev. Lett. 103, 117201 (2009).

[12] Tudosa, I., Stamm, C., Kashuba, A.B., King, F.,  Siegmann, H.C., Stöhr, J., Ju, G.,  Lu B. & Weller,D.  . Nature 428, 831-833 (2004)

[14] Steil, D.,  Alebrand, S.,  Hassdenteufel, A.,  Cinchetti, M., & Aeschlimann, M.. Phys. Rev. B 84, 224408 (2011)

[16]  Alebrand, S., Hassdenteufel, A. , Steil, D., Cinchetti, M. & Aeschlimann, M. Phys. Rev. B 85, 092401 (2012)

[17] Hansen, P. ,  Clausen, C. ,  Much, G. ,  Rosenkranz, M. & Witter, K.  J. Appl. Phys. **66**, 756 (1989)

[18] Luborsky, F.E.. Mat. Res. Soc. Symp. Proc. Vol 80. 375-384(1987)

[19] Gottwald, M., Hehn, M., Lacour, D., Hauet, T., Montaigne F., Mangin S., Fischer, P., Im M.-Y., Berger, A. , Phys. Rev. B 85, 064403 (2012)

[20] Gottwald, M., Hehn, M., Montaigne, F., Lacour, D., Lengaigne, G., Suire, S., Mangin, S., J. Appl. Phys 111 083904 (2012)

[21] Aeschlimann, M., Vaterlaus, A., Lutz, M., Stampanoni, M. & Meier, F. Ultrafast thermomagnetic writing processes in rareearth transitionmetal thin Films. J. Appl. Phys. 67, 4438 (1990); doi: 10.1063/1.344924

[22] Jiang, X Li Gao, Jonathan Z. Sun, and Stuart S. P. Parkin,. Phys. Rev. Lett. 97, 217202 (2006)

[23] Dalla Longa, F. , Kohlhepp, J. T. , de Jonge, W. J. M. , & Koopmans, B.  . Phys. Rev. B 75, 224431  (2007)

[24] A. Kirilyuk, A.V. Kimel and T. Raising Review of Modern Physics Vol 80 2012